\documentclass[10pt,letterpaper,twoside,final,twocolumn,journal]{IEEEtran}
\usepackage[english]{babel}
\addto\captionsenglish{}
\usepackage{setspace}
\usepackage{fixltx2e}
\usepackage{graphicx}
\usepackage[caption=false]{subfig}
\usepackage{algorithm}
\usepackage{algorithmic}
\usepackage{epsfig}
\usepackage{epstopdf}
\usepackage[cmex10]{amsmath}
\usepackage{cite}
\usepackage{amssymb}
\usepackage{amsthm}
\usepackage{color}
\usepackage{multirow}
\usepackage{array}

\newcommand{\abs}[1]{\left| #1 \right|} 

\let\baraccent=\= 
\renewcommand{\=}[1]{\stackrel{#1}{=}} 

\usepackage[color]{changebar}
\usepackage{xcolor}
\usepackage{tabularx}
\usepackage{chngcntr}

\usepackage{epstopdf}
\DeclareGraphicsExtensions{.eps,.pdf,.png,.jpg}

\usepackage{mathtools}

\theoremstyle{definition}

\theoremstyle{definition}

\PassOptionsToPackage{hyphens}{url}\usepackage[breaklinks=true, colorlinks=true, citecolor=blue, bookmarksopen=true]{hyperref}

\begin{document}
\title{
{Quantum-aided Multi-Objective Routing Optimization Using Back-Tracing-Aided Dynamic Programming}}
\author{
Dimitrios~Alanis,~\IEEEmembership{Student~Member,~IEEE,}
Panagiotis~Botsinis,~\IEEEmembership{Member,~IEEE,} 
Zunaira~Babar,
Hung Viet Nguyen,~\IEEEmembership{Member,~IEEE,}
Daryus Chandra,~\IEEEmembership{Student~Member,~IEEE,} 
Soon~Xin~Ng,~\IEEEmembership{Senior~Member,~IEEE,}  and~Lajos~Hanzo,~\IEEEmembership{Fellow,~IEEE}\vspace*{-0.5cm}%
\thanks{The authors are with the School of Electronics and Computer Science, University of Southampton, Southampton, SO17 1BJ, UK (email: \{da1d16,~pb1y14,~zb2g10,~hvn08r,~dc2n14,~sxn,~lh\}@ecs.soton.ac.uk).}
\thanks{The financial support of the EPSRC under the grant EP/L018659/1, that of the European Research Council, Advanced Fellow Grant and that of the Royal Society's Wolfson Research Merit Award is gratefully acknowledged. Additionally, the authors acknowledge the use of the IRIDIS High Performance Computing Facility, and associated support services at the University of Southampton, in the completion of this work. he research data of this
paper can be found at \url{https://doi.org/10.5258/SOTON/D0479}.}
}
\markboth{IEEE Transactions on Vehicular Technology 2018}{D.~Alanis \emph{et al.}:Quantum-aided Multi-objective Routing Optimization using Back-tracing-enabled Dynamic Programming}
\maketitle
\begin{abstract}
Pareto optimality is capable of striking the optimal trade-off amongst the diverse conflicting QoS requirements of routing in wireless multihop networks. However, this comes at the cost of increased complexity owing to searching through the extended multi-objective search-space. We will demonstrate that the powerful quantum-assisted dynamic programming optimization framework is capable of circumventing this problem. In this context, the so-called Evolutionary Quantum Pareto Optimization~(EQPO) algorithm has been proposed, which is capable of identifying most of the optimal routes at a near-polynomial complexity versus the number of nodes. As a benefit, we improve both the the EQPO algorithm by introducing a back-tracing process. We also demonstrate that the improved algorithm, namely the Back-Tracing-Aided EQPO~(BTA-EQPO) algorithm, imposes a negligible complexity overhead, while substantially improving our performance metrics, namely the relative frequency of finding all Pareto-optimal solutions and the probability that the Pareto-optimal solutions are indeed part of the optimal Pareto front.
\end{abstract}

\section{Introduction}
\IEEEPARstart{R}{outing} optimization in \emph{Wireless Multihop Networks}~(WMHN) has to strike a trade-off among diverse and often conflicting \emph{Quality-of-Service}~(QoS) requirements\cite{fei2016asurvey}. For this reason several metrics have been advocated, such as the \emph{Network Lifetime}~(NL) \cite{tashtarian2015onmaximizing} or the \emph{Network Utility}~(NU) \cite{zhou2016joint}, which are \emph{single-objective} aggregate functions of multiple QoS requirements. However, these single-objective metrics may not be giving justice to all design objectives. This problem can be circumvented by employing the concept of Pareto optimality \cite{liao2015amultiutility,deb2005mo}. This comes at the cost of increased complexity imposed by the extended search-space, which can be in turn circumvented by utilizing the powerful optimization framework of quantum computing \cite{nielsen2010quantum}.

In this context, several contributions on quantum-aided multi-objective routing exist in the literature \cite{alanis2014ndqo, alanis2015ndqio, alanis2016modqo, alanis2017eqpo}. To elaborate further, the so-called \emph{Non-Dominated Quantum Optimization}~(NDQO) and the \emph{Non-Dominated Quantum Iterative Optimization}~(NDQO) algorithms have been proposed in \cite{alanis2014ndqo} and \cite{alanis2015ndqio}, respectively, relying on full-search-based database exploration. As an intermediate step, the so-called \emph{Non-Dominated Quantum Optimization}~(MODQO) algorithm of \cite{alanis2016modqo} exploited the database correlations emerging from the formation of Pareto-optimal route-combinations for efficiently reducing the database size, thus achieving a further complexity reduction. The database correlation has been exploited in \cite{alanis2017eqpo}, where the \emph{Evolutionary Quantum Pareto Optimization}~(EQPO) algorithm has been introduced. More explicitly, the EQPO algorithm, which is a feed-forward-style algorithm, achieved a further complexity reduction by exploiting the potential correlations among the individual links constituting Pareto-optimal routes. Nevertheless, this complexity reduction comes at the price of reduced heuristic accuracy.
Against this background our contributions are summarized as follows:
\textit{
\begin{enumerate}
\item[\emph{1)}] We propose an improved version of the EQPO, namely the Back-Tracing-Aided EQPO~(BTA-EQPO) algorithm, by introducing novel Back-Tracing Processes (BTPs) by extending the quantum-aided dynamic programming framework of \cite{alanis2017eqpo}.
\item[\emph{2)}] We demonstrate that the BPTs impose an insignificant complexity overhead, when compared to the complexity imposed by the EQPO algorithm, hence the BTA-EQPO imposes the same order of complexity as its predecessor, namely the EQPO.
\item[3)] We also demonstrate that the BTA-EQPO algorithm's resultant error floor is an order of magnitude below that of its predecessor. 
\end{enumerate}
}
The rest of this paper is organized as follows. In Section~II, we will present the network topology considered. In Section~III, we will elaborate on the novel back-tracing process of the BTE-EQPO algorithm. We will then evaluate its performance versus complexity in Section IV.

\section{Network Specifications}
\begin{figure}[htb]
\centering\vspace{-0.3cm}
\includegraphics[width=0.6\linewidth]{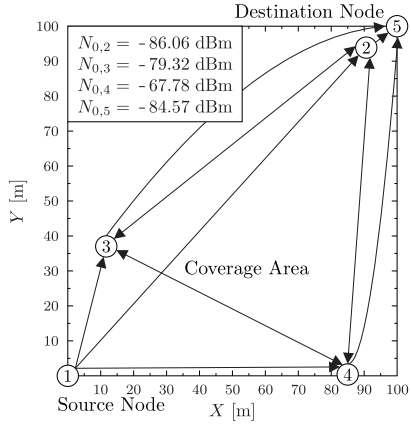}
\caption{Exemplified WMHN topology associated with $N_\text{nodes}=5$ nodes \cite{alanis2017eqpo}.\label{fig:network-topology}}
\end{figure}

We have adopted the WMHN model considered in \cite{alanis2014ndqo, alanis2015ndqio, alanis2017eqpo}, where the \emph{Source Node}~(SN) and the \emph{Destination Node}~(DN) are located at the opposite corners of a $(100\times 100)$~m$^2$ square block. By contrast, the \emph{Relay Nodes}~(RNs) are mobile, having locations that are uniformly distributed within this square block. We also assume that the DN acts as a cluster-head,  which has access to a universal quantum computer. Each node experiences random interference power, relying on a normal distribution with its mean set to -90~dBm and its standard deviation to 10~dB. An example of the network topology consisting of $N_\text{nodes}=5$ nodes is shown in Fig.~\ref{fig:network-topology}. 
As for our optimization metrics, we have jointly considered the routes' end-to-end delay $D$, their total \emph{Bit Error Ratio}~(BER) $P_e$ as well as their total power dissipation $L$ in a similar fashion to \cite{alanis2014ndqo, alanis2015ndqio, alanis2017eqpo}. More specifically, we have considered \emph{Quadrature Phase Shift Keying}~(QPSK) transmissions in an uncorrelated Rayleigh fading environment, where the packet forwarding has been carried out using the \emph{Decode-and-Forward}~(DF) scheme \cite{yang2015isthe}. Consequently, the route's overall BER $P_e(x)$ can be calculated using the following recursive formula \cite{alanis2014ndqo}:
\begin{equation}\label{eq:ber_rec}
P_{e,tot}=P_{e,1}+P_{e,2}-2P_{e,1}P_{e,2},
\end{equation}
where $P_{e,tot}$ corresponds to the output BER  of a two-stage \emph{Binary Symmetric Channel} (BSC) \cite{alanis2014ndqo} with $P_{e,1}$ and $P_{e,2}$ representing the individual BER of the first and the second stage, respectively. Additionally, the route's end-to-end delay $D$ is quantified in terms of the number of hops composing the route, while the total power dissipation $L$ is determined by the sum of the path-losses of each individual link $L_{ij}$ of the route. Explicitly, each link between the $i$-th and the $j$-th nodes exhibits path-losses quantified in dB as follows \cite{alanis2015ndqio}:
\begin{equation}
L^{dB}_{ij} = 10\alpha\log_{10}\left(\frac{4 \pi d_{ij}}{\lambda_c}\right),
\end{equation}
where $\alpha$ corresponds to the path-loss exponent, which is set to $\alpha=3$, $d_{ij}$ denotes the Euclidean distance between the $i$-th and the $j$-th, while $\lambda_c$ is the carrier's wavelength set to $\lambda_c=0.125$~m. Therefore, the \emph{Utility Vector}~(UV) $\mathbf{f}(x)$ of the $x$-th route can be expressed as follows:
\begin{equation}\label{eq:uv}
\mathbf{f}(x)=\left[P_e(x),L(x),D(x)\right].
\end{equation}
The concept of Pareto optimality~\cite{deb2005mo} has been adopted for evaluating the fitness of the UVs. In a nutshell, a specific route $x_1$ dominates another route $x_2$, i.e. we have $\mathbf{f}(x_1)\succ\mathbf{f}(x_2)$, if all the individual metrics of $\mathbf{f}(x_1)$ are lower than the respective components of $\mathbf{f}(x_2)$. Based on this principle, a route is considered to be Pareto optimal, if there are no other routes dominating it. Note that our ultimate goal is to identify the entire set of Pareto optimal routes, which jointly constitute the so-called \emph{Optimal Pareto Front}~(OPF) \cite{alanis2014ndqo}.
 
\section{Back-Tracing-Aided Quantum Pareto Optimization}
The BTA-EQPO algorithm, which is presented in Alg.~\ref{alg:bte-eqpo}, is constituted by three distinct parts: a stage of the single-objective optimization followed by the so-called \emph{Single-Objective Back-Tracing Process}~(SO-BTP), a stage of the multi-objective optimization process in a similar fashion to the EQPO algorithm and a stage invoking a  \emph{Multi-Objective Back-Tracing Process}~(MO-BTP).
\begin{algorithm}[htb]
\caption{Back-Tracing-Aided Evolutionary Quantum Pareto Optimization (BTA-EQPO) Algorithm\label{alg:bte-eqpo}.}
\begin{algorithmic}[1]
\STATE Set  $S^\text{OPF}_{(i)}\leftarrow \varnothing$ and $S^\text{surv}_{(i)}\leftarrow \varnothing$ $\forall i \in\{0,..,N_\text{nodes}-1\}$.
\STATE Determine the optimal routes $S^\text{opt}$ based on each individual objective based on the optimal framework presented in \cite[Sec. III]{alanis2017eqpo} and store  accordingly the optimal routes visited in $S^\text{OPF}_{(i)}$, where $i$ is the number of RNs constituting the visited route.
\STATE For each route in $S^\text{opt}$ perform SO-BTP based on Fig.~\ref{fig:bt-fig} and store  accordingly the surviving routes visited to the set $S^\text{surv}_{(i)}$, where $i$ is the number of RNs constituting the visited route.
\STATE Set $S^{\text{gen}}_{(0)} \leftarrow \{SN\rightarrow DN\}$,  $S^\text{OPF}_{(0)}\leftarrow S^{\text{gen}}_{(0)}$, $S^\text{surv}_{(0)}\leftarrow S^{\text{gen}}_{(0)}$, $i\leftarrow 0$.
\REPEAT
\STATE Set $i\leftarrow i+1$.
\STATE Generate the set of routes $S^{\text{gen}}_{(i)}$ from the set $S^\text{surv}_{(i-1)}$ by appropriately inserting a single RN between two intermediate nodes.
\STATE Set $S^{\text{gen}}_{(i)}\leftarrow S^{\text{gen}}_{(i)}\cup S^\text{OPF}_{(i-1)}$. 
\STATE Invoke the P-NDQIO algorithm of \cite[Alg.~2]{alanis2017eqpo} in the set $S^{\text{gen}}_{(i)}$ and initialize the identified OPF to $S ^\text{OPF}_{(i)}\leftarrow S ^\text{OPF}_{(i-1)}\cup S^\text{OPF}_{(i)}$.
\STATE Set $S^\text{surv}_{(i)}\leftarrow \left(S^\text{OPF}_{(i)} - S^\text{OPF}_{(i-1)}\right)\cup S^\text{surv}_{(i)}$. 
\UNTIL{$\abs{S^\text{surv}_{(i)}}=0$ \OR $i=N_\text{nodes}-2$}
\STATE Set $i\leftarrow i+1$.
\STATE For each route in $S^\text{OPF}_{(i)}$ perform MO-BTP  for $n$ trellis-stages based on Fig.~\ref{fig:bt-fig} and store the surviving routes visited in $S^\text{gen}_{(i)}$.
\STATE Invoke the P-NDQIO algorithm of \cite[Alg.~2]{alanis2017eqpo} in the set $S^{\text{gen}}_{(i)}\leftarrow S^{\text{gen}}_{(i)} \cup S ^\text{OPF}_{(i-1)}$ and initialize the identified OPF to $S ^\text{OPF}_{(i)}\leftarrow S ^\text{OPF}_{(i-1)}$.
\STATE Export the OPF $S ^\text{OPF}_{(i)}$ and terminate.
\end{algorithmic}
\end{algorithm}

As far as the first stage is concerned, we first invoke in Step~2 of Alg.~\ref{alg:bte-eqpo} single-objective dynamic programming based optimization utilizing the optimal dynamic programming framework of \cite[Sec. III]{alanis2017eqpo} for the sake of identifying the optimal routes $S^\text{opt}$ in terms of each individual objective. These routes will also be Pareto-optimal \cite{deb2005mo}, when jointly optimizing the UV of Eq.~(\ref{eq:uv}). Therefore, we will appropriately initialize the set  $\{S^\text{OPF}_{(i)}\}_{i=0}^{N_\text{nodes}-2}$ of Pareto-optimal routes to the set $S^\text{opt}$ based on the trellis-stage index $i$, during which they were identified.  For instance, the optimal route $1\rightarrow 2 \rightarrow 3 \rightarrow 4 \rightarrow 5$ will be appended to $S^\text{OPF}_{(3)}$, since it consists of 3 RNs and thus it was identified at the second trellis-stage. We have opted for this optimal framework, since it guarantees the detection of these globally optimal routes, while it imposes a complexity\footnote{We quantify the complexity in terms of the number of dominance comparisons; a single dominance comparison is defined as a single \emph{Cost Function Evaluation}~(CFE). We further distinguish the complexity into two domains: the \emph{parallel complexity}\cite{alanis2017eqpo}, which takes into account the beneficial hardware parallelism exploited by the NDQIO-based algorithms, and the \emph{sequential complexity}\cite{alanis2017eqpo}, which neglects the benefits of hardware parallelism and it is simply quantified in terms of the number of Pareto-dominance comparisons. In our application we have utilized quantum Pareto-dominance comparison operators that are identical to those of \cite{alanis2015ndqio}. Consequently, assuming a total of $a$ reference routes and $k$ optimization objectives, a single activation of this quantum dominance operator results in a parallel and a sequential complexity of $1/k$ and $a$ \emph{Cost Function Evaluations}~(CFEs), respectively.} on the order of $O(N_\text{nodes}^3)$. Explicitly, there exist precisely $N_\text{nodes}$ surviving routes at each trellis-stage, thus a total of $N_\text{nodes}^2$ comparisons are required per trellis-stage, while a total of $O(N_\text{nodes})$ trellis stages are processed. 

\begin{figure}[htb]
\centering
\includegraphics[width=0.85\linewidth]{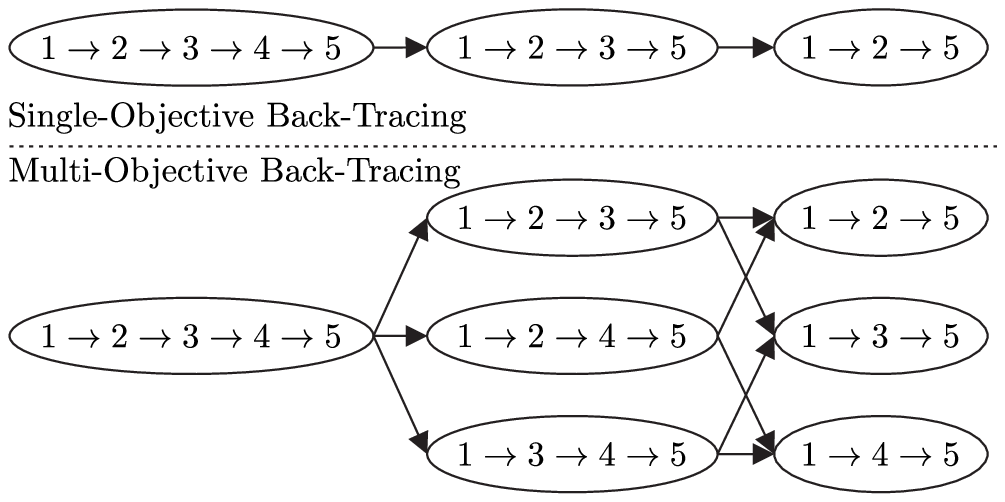}
\caption{Single- versus multi-objective back-tracing.\label{fig:bt-fig}}
\end{figure}

Subsequently, the SO-BTP is activated in Step~3 of Alg.~\ref{alg:bte-eqpo} for each of the globally optimal routes identified by the optimization process in Step~2 of Alg.~\ref{alg:bte-eqpo}. During this process, starting from a single optimal route we successively trace back to the direct route by removing the last RN of the route, as portrayed in the upper sub-figure of Fig.~\ref{fig:bt-fig}. We conceived utilized this specific strategy, since the routes of a specific trellis stage are generated by appropriately inserting an RN between the last RN and the DN at each of the surviving routes of the previous trellis-stage. Additionally, the surviving routes w.r.t. an individual objective will be also classified as surviving \cite{deb2005mo}, when we jointly optimize the entire set of objectives, since their sub-routes will remain non-dominated by any other route or sub-route. Using this observation, we will appropriately initialize the set $\{S^\text{surv}_{(i)}\}_{i=0}^{N_\text{nodes}-2}$ of surviving routes to the specific routes visited during each of the SO-BTPs in a similar fashion to the initialization of $\{S^\text{OPF}_{(i)}\}_{i=0}^{N_\text{nodes}-2}$. 

After the initialization of the surviving routes, a multi-objective optimization process similar to that of the EQPO algorithm of \cite[Alg.~1]{alanis2017eqpo} is activated in Steps~5-11 of Alg~\ref{alg:bte-eqpo}. Their main difference is that both the set of surviving and Pareto optimal routes have been initialized by the SO-BTP, as highlighted in Steps~9 and 10 of Alg.~\ref{alg:bte-eqpo}. Naturally, the initialization of the surviving routes expands the search-space, hence rendering the BTA-EQPO capable of identifying a more diverse set of Pareto optimal routes. This search-overhead imposed by the additionally generated routes is on the order of $O(N_\text{nodes})$ extra cost-function evaluations, when using a similar approach to that of \cite{alanis2017eqpo}. Since the number of generated routes excluding this overhead at the $i$-th trellis-stage is on the order of $O(N_\text{OPF}N_\text{nodes}i)$ with $N_\text{OPF}$ representing the number of Pareto optimal routes, we may deem this overhead to be low.  Quantitatively, the second step imposes the same order of complexity as the EQPO algorithm, whose parallel and sequential complexity were shown to be on the order of $O(N_\text{OPF}^{3/2}N_\text{nodes}^2)$ and $O(N_\text{OPF}^{5/2}N_\text{nodes}^2)$, respectively. Naturally, the complexity order of the fist stage can be considered as negligible compared to that of the second stage.

Finally, the third stage in Steps~13-14 of Alg.~\ref{alg:bte-eqpo} is activated, which invokes the MO-BTP for $n$ trellis stages and it is invoked for each of the hitherto identified OPF routes. To further aid its exposition, its employment is visually portrayed in the bottom sub-figure of Fig.~\ref{fig:bt-fig}. During the MO-BTP, the inverse  of Step~7 of Alg.~\ref{alg:bte-eqpo} is carried out, i.e. we move to the previous trellis stage by removing a single RN from the route examined. For instance, observe in Fig.~\ref{fig:bt-fig} that invoking the MO-BTP for the Pareto optimal route $1\rightarrow 2 \rightarrow 3 \rightarrow 4 \rightarrow 5$ results in visiting the routes $1\rightarrow 2 \rightarrow 3 \rightarrow 5$, $1\rightarrow 2 \rightarrow 4 \rightarrow 5$ and $1\rightarrow 3 \rightarrow 4 \rightarrow 5$, when back-tracing  for $n=1$ trellis stage, and the routes $1\rightarrow 2\rightarrow 5$, $1\rightarrow 3\rightarrow 5$ as well as $1\rightarrow 4\rightarrow 5$, when back-tracing for $n=2$ trellis stages. During this process, we keep track of the visited routes of the MO-BTP, storing them while we reach the final set $S^\text{gen}_{(i)}$ of generated routes. We then invoke the \emph{Preinitialized NDQIO}~(P-NDQIO) algorithm \cite[Alg.~2]{alanis2017eqpo} with its OPF initialized to the hitherto identified OPF emanating from the second stage for the sake of finding any further Pareto optimal routes. The complexity order of the P-NDQIO algorithm is proportional to $O(\sqrt{N})$ \cite{alanis2017eqpo}. We have chosen to optimize the routes over the entire database, since offers a beneficial complexity reduction against performing the optimization for each backward trellis transition, since we have $\sqrt{\sum_{i}n_i}<\sum_{i}\sqrt{n_i}$. 

Last but not least, let us quantify the extra complexity imposed by this process. The total number of generated routes as a function of the number $n$ of backward trellis transitions can be readily shown to be on the order of $O(N_\text{OPF}N_\text{nodes}^n)$. Consequently, the parallel and sequential complexities imposed by the P-NDQIO algorithms of the MO-BTP may be shown to be on the orders of $O(N^{3/2}_\text{OPF}N_\text{nodes}^{n/2})$ and $O(N^{5/2}_\text{OPF}N_\text{nodes}^{n/2})$, respectively. Hence, the total complexity imposed by the BTE-EQPO can be shown to be:
\begin{align}
L^P_\text{BTA-EQPO}&=O\left[N^{3/2}_\text{OPF}\left(N_\text{nodes}^{2}+N_\text{nodes}^{n/2}\right)\right],\label{eq:L-BTE-EQPO-P}\\
L^S_\text{BTA-EQPO}&=O\left[N^{5/2}_\text{OPF}\left(N_\text{nodes}^{2}+N_\text{nodes}^{n/2}\right)\right].
\end{align}
Hence, the MO-BTP will dominate the complexity orders, when having more than $n=4$ backward-trellis steps. Let us now proceed by examining the performance versus complexity trade-off of the BTA-EQPO algorithm.
\begin{figure}[t]
	\subfloat[][Parallel Complexity\label{fig:complexity-p}]{\includegraphics[width=\linewidth]{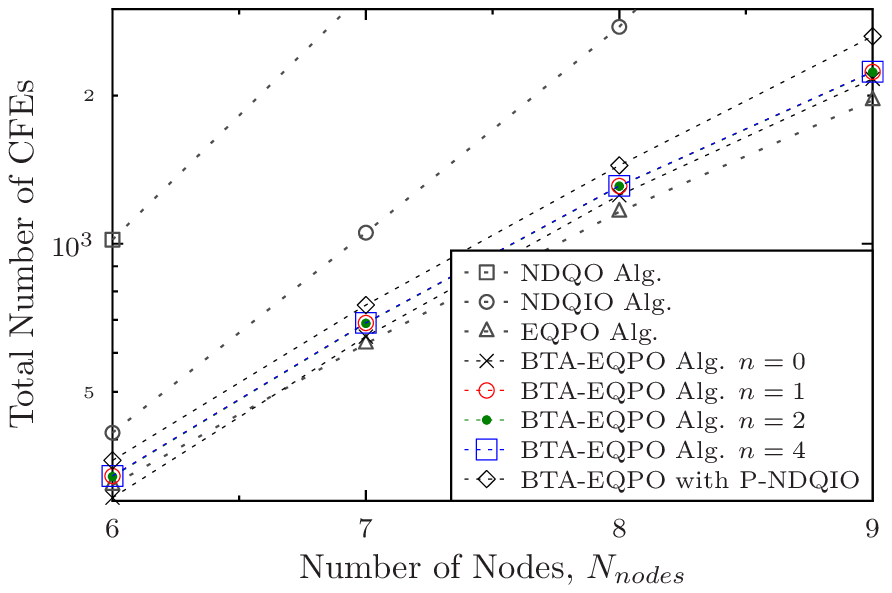}}\hfill
\vspace*{-0.4cm}
  	\subfloat[][Sequential Complexity\label{fig:complexity-s}]{\includegraphics[width=\linewidth]{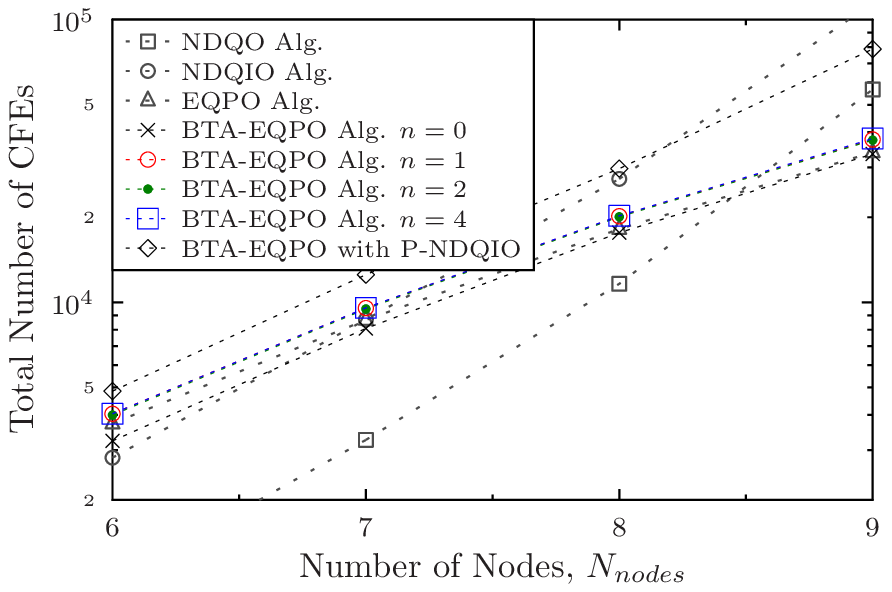}}\hfill
\caption{Parallel~(a) and sequential~(b) complexities of the BTE-EQPO algorithm compared to those of the EQPO, NDQIO and NDQIO algorithms. The results have been averaged over 10$^8$ runs.\label{fig:complexity}}
\end{figure}
\section{Performance versus Complexity}
In this section we will provide some further insights concerning BTA-EQPO algorithm's performance versus complexity and compare it to the existing quantum assisted algorithms, namely the EQPO~\cite{alanis2017eqpo}, NDQIO~\cite{alanis2015ndqio} and NDQO~\cite{alanis2014ndqo} algorithms. We will first examine the average complexity imposed by the aforementioned algorithms as a function of the number $N_\text{nodes}$ of nodes constituting the WMHN. 
In addition to the aforementioned algorithms we investigate a hybrid algorithm, which uses the first two stages of the BTA-EQPO, while the third stage is replaced by a full-database search carried out by the P-NDQIO algorithm~\cite{alanis2017eqpo}. The latter will be referred to as ``BTA-EQPO with P-NDQIO'' and it is used as the upper bound of the complexity imposed by MO-BTP, when we have $n=N_\text{nodes}-1$.

The average parallel and sequential complexities are shown in Figs.~\ref{fig:complexity-p} and \ref{fig:complexity-s}. In these figures we vary the number $n$ of backward trellis stages in the range of $\{0,1,2,4\}$. Note that for $n=0$ only the SO-BTP is active, while for $n=4$ the MO-BTP complexity orders match those of the BTA-EQPO algorithm's second stage. Observe in both figures that both the parallel and the sequential complexity imposed by the BTA-EQPO algorithm approach that of the EQPO algorithm, hence verifying our theoretical analysis of Sec.~III, where we proved that the extra complexity imposed both by the SO-BTP and by the MO-BTP is significantly lower than the complexity of BTA-EQPO algorithm's second stage. Furhtermore, observe in Fig.~\ref{fig:complexity-p} that the BTA-EQPO algorithm imposes almost the same parallel complexity as BTA-EQPO with P-NDQIO algorithm. However, a a factor of two sequential complexity increase is observed in Fig.~\ref{fig:complexity-s} for 9-node WMHNs. This is because the square root of the total number of routes is close to that of the routes created by the MO-BTP for the WMHN sizes we investigated; however, for larger WMHNs we expect much higher complexity reduction for our BTA-EQPO algorithm. Additionally, both a parallel and a sequential complexity reduction is achieved against the NDQIO algorithm, which almost is a high as an order of magnitude for 9-node WMHNs.
\begin{figure*}[t]
\centering
  	\subfloat[][\label{fig:pd-p}]{\includegraphics[width=0.5\linewidth]{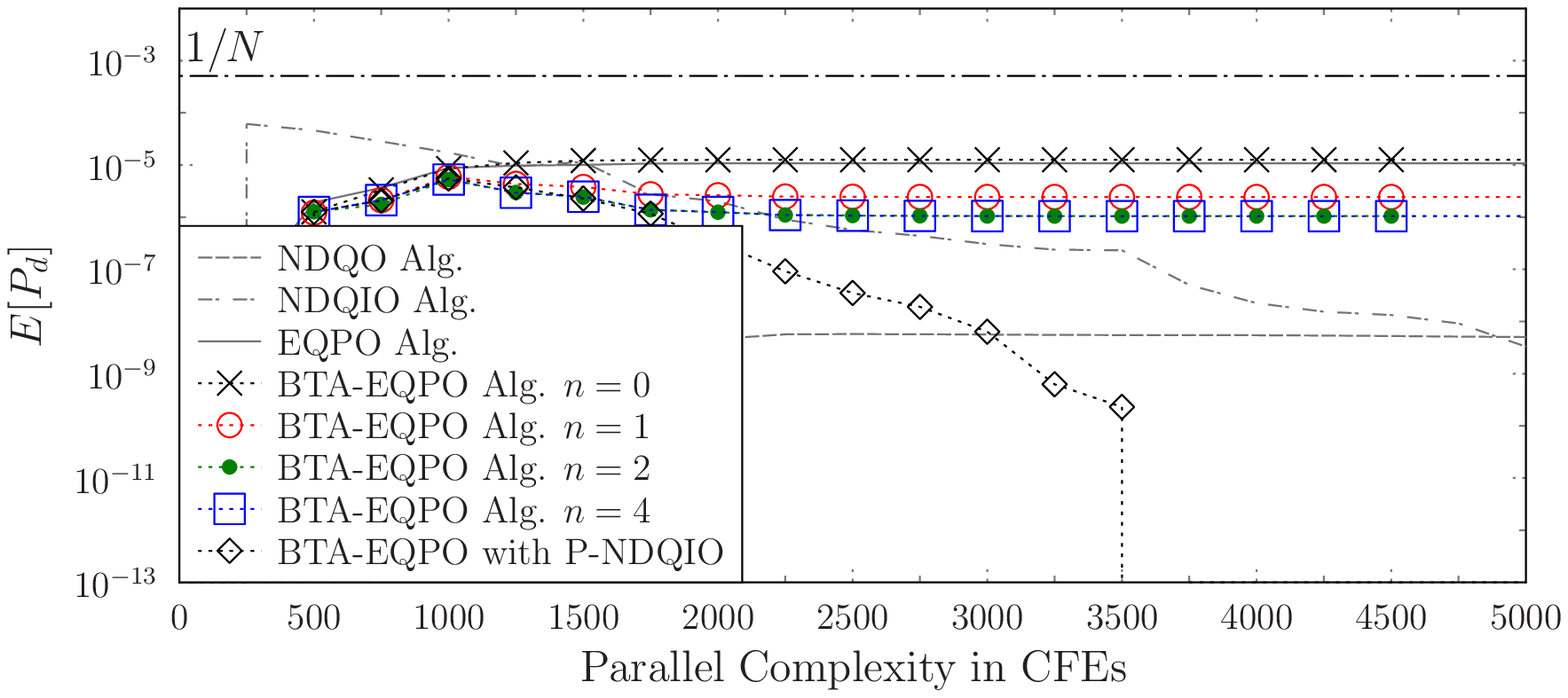}}\hfill
  	\subfloat[][\label{fig:pd-s}]{\includegraphics[width=0.5\linewidth]{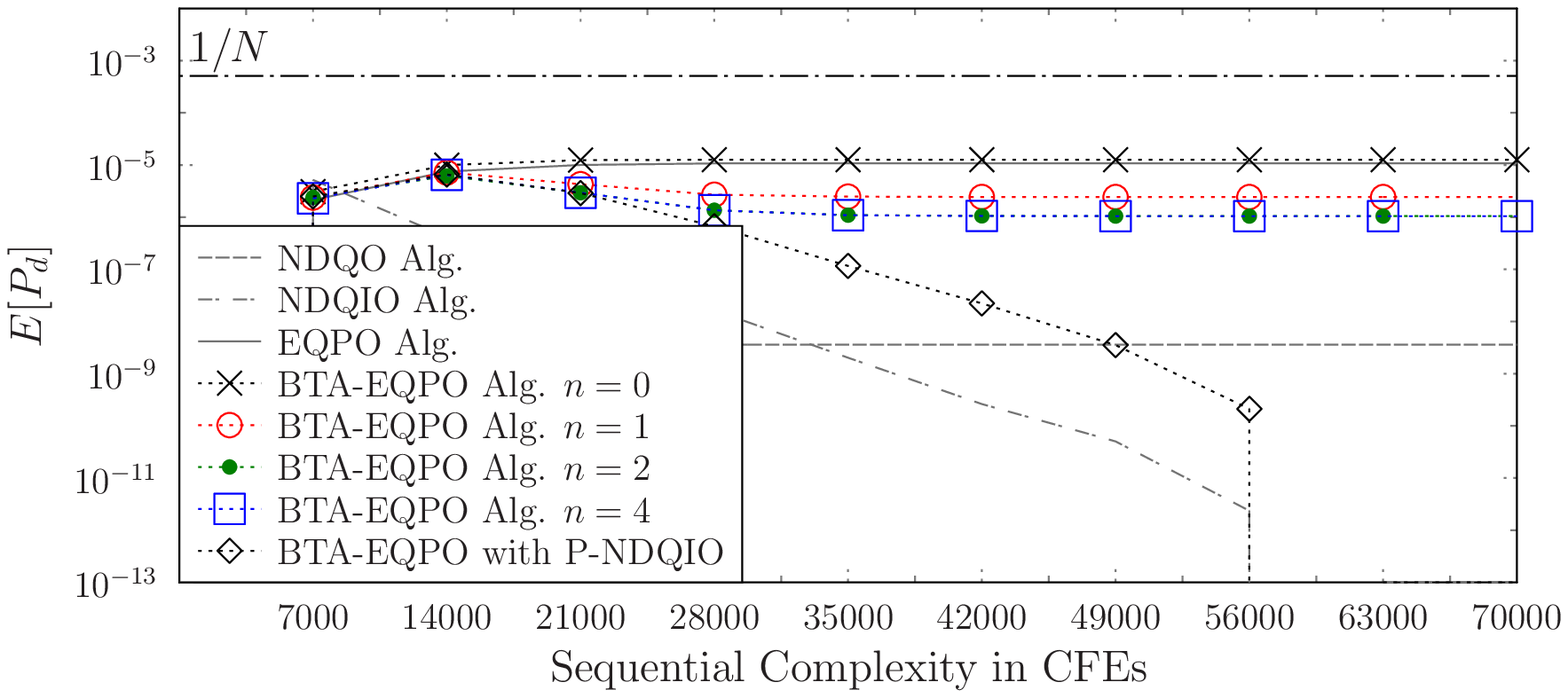}}\hfill
  	\subfloat[][\label{fig:c-p}]{\includegraphics[width=0.5\linewidth]{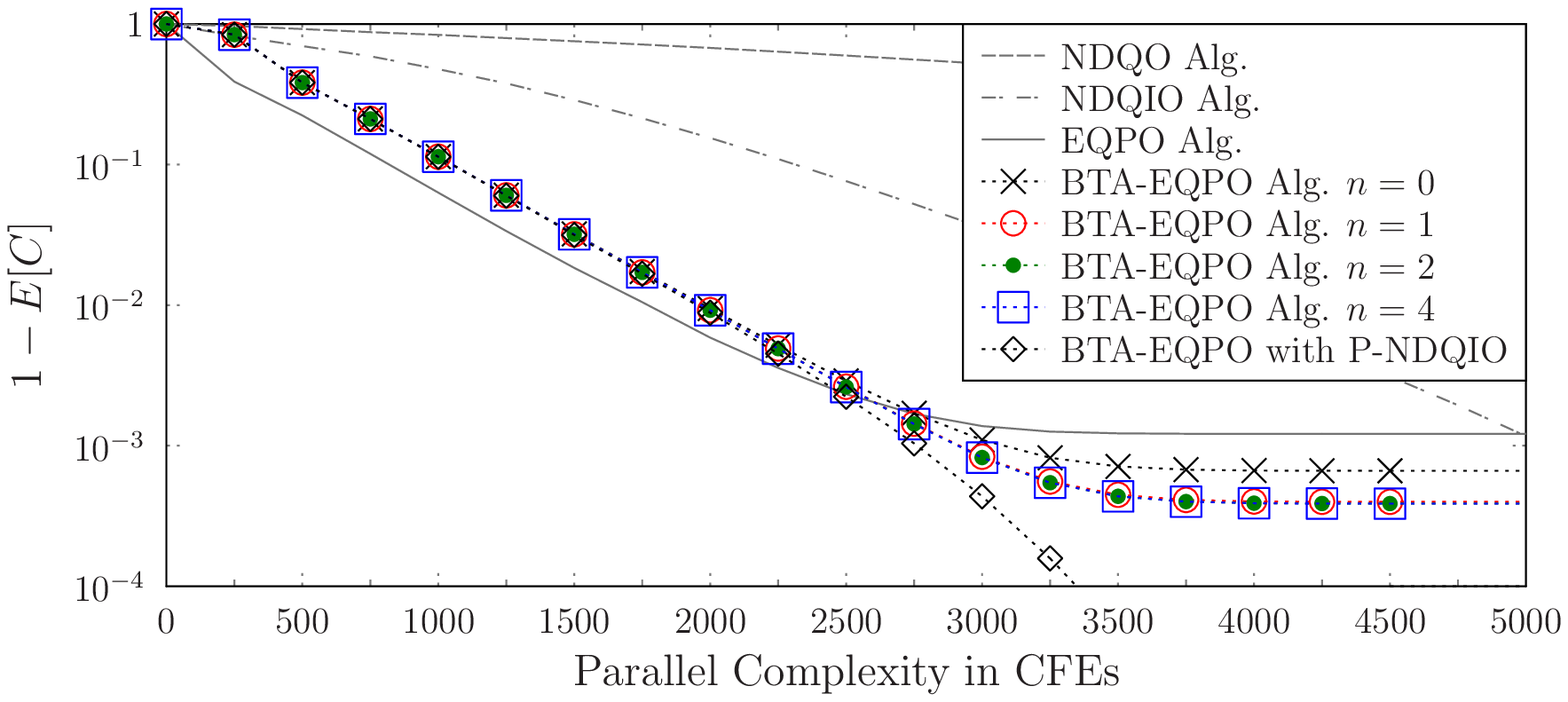}}\hfill
  	\subfloat[][\label{fig:c-s}]{\includegraphics[width=0.5\linewidth]{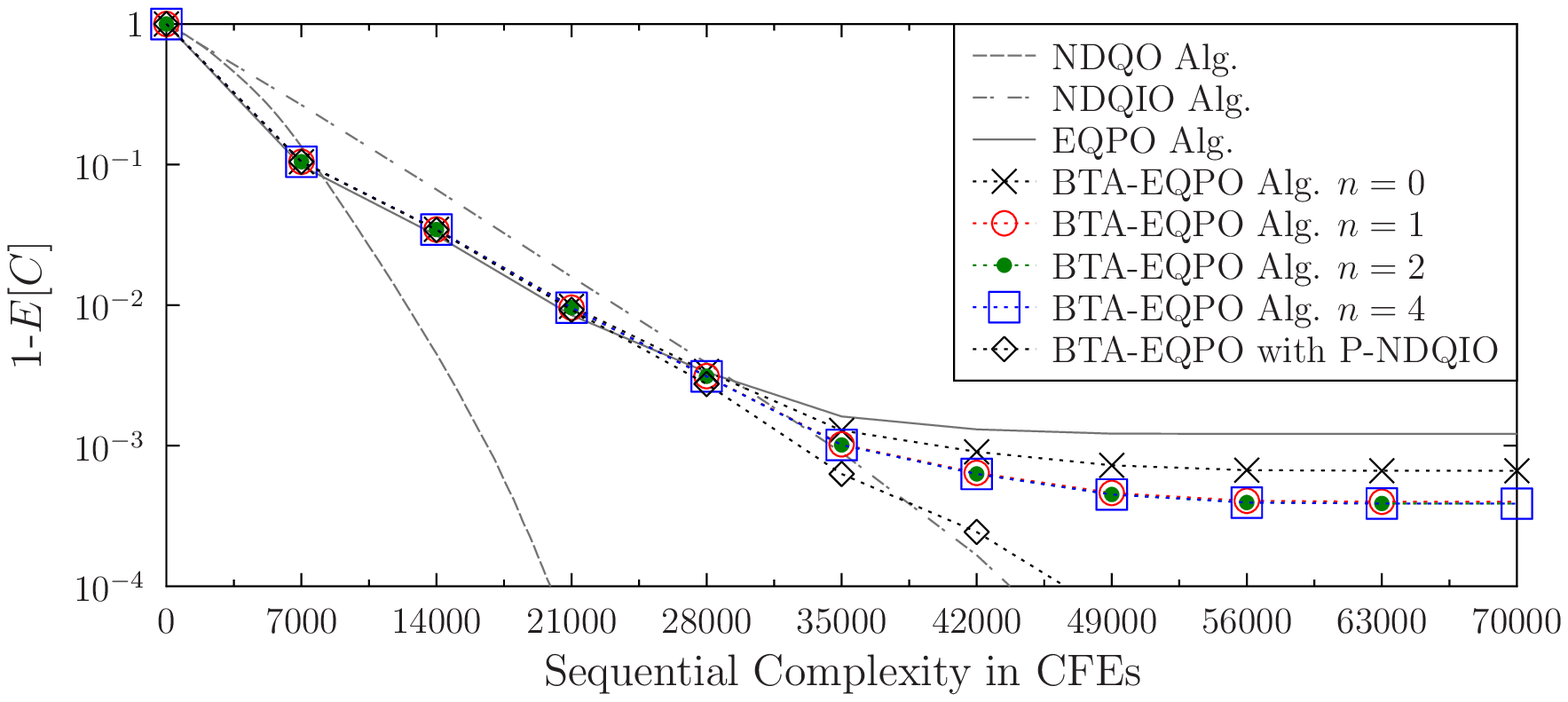}}\hfill
    \caption{Performance versus complexity in terms of the average Pareto distance $E[P_d]$~(a,c) and the average Pareto completion $E[C]$ for 8-node WMHNs. In (b) and (d) the value $1-E[C]$ is portrayed facilitating observation of the differences among the algorithms. The results have been averaged over 10$^8$ runs.\label{fig:accuracy}}
\end{figure*}

Continuing with the BTA-EQPO algorithm's performance evaluation, we will utilize two metrics: the average \emph{Pareto distance} $E[P_d]$ \cite{alanis2014ndqo}, which is defined as the probability of a route identified as Pareto optimal being truly Pareto optimal, and the average \emph{Pareto completion} $E[C]$ \cite{alanis2014ndqo}, defined as the average fraction of the true OPF being identified by a heuristic method. Naturally, for $E[P_d]=0$ the identified OPF exclusively consists of true Pareto optimal routes, while for $E[C]=1$ the entire true OPF has been identified. The average Pareto distance $E[P_d]$ is shown in Figs.~\ref{fig:pd-p} and \ref{fig:pd-s} as a function of the  parallel and sequential complexity invested, respectively. Observe in these figures that the BTA-EQPO algorithm associated with $n=0$, i.e. with the particular case where SO-BTP is active, has a similar performance to that of the EQPO algorithm~\cite{alanis2017eqpo}. However, the beneficial effects of MO-BTP are visible even for $n=1$, where $E[P_d]$ is reduced by a factor of 5 after 2,200 and 28,000 CFEs in the parallel and sequential complexity domains, respectively, when compared to the EQPO algorithm, where the latter is portrayed with the aid of the gray solid lines. This improvement is further enhanced for $n=2$ and $n=4$, where $E[P_d]$ is improved by an order of magnitude compared to that of the EQPO algorithm. Additionally, observe in Figs.~\ref{fig:pd-p} and \ref{fig:pd-s} that beyond $n=2$ the BTP-EQPO algorithm exhibits an error floor formation, hence rendering the application of further backward-trellis steps redundant. As for the full-search-based methods, observe in Fig.~\ref{fig:pd-p} that the BTA-EQPO with P-NDQIO algorithm becomes more efficient than both the NDQO and the NDQIO algorithms beyond a parallel complexity of 3,000 CFEs, while its $E[P_d]$ decays to infinitesimally low levels beyond 3,500 CFEs. This trend is also present in Fig.~\ref{fig:pd-s}; however, observe that the NDQIO algorithm is more efficient than the BTA-EQPO with P-NDQIO algorithm. However, we expect this trend to change following that of Fig.~\ref{fig:pd-p} as the number of nodes increases, where the BTA-EQPO with P-NDQIO algorithm offers a substantial sequential complexity reduction compared to the NDQIO algorithm.

As far as the average Pareto completion is concerned, observe in Figs.~\ref{fig:c-p} and \ref{fig:c-s} that the BTA-EQPO algorithm associated with $n=0$ succeeds in identifying a larger fraction of the OPF by improving the complementary Pareto completion metric by a factor of 3. This happens at a parallel and a sequential complexity of 3,500 and 49,000 CFEs, respectively, thus explicitly demonstrating the benefit of the SO-BTP. When the MO-BTP is activated, this metric is further reduced, exhibiting of an order of magnitude total improvement over EQPO algorithm. Additionally, we can observe that this metric is slightly improved, as the number $n$ of backward-trellis steps increases. Explicitly, the Pareto Completion error floor exhibited stems from the BTA-EQPO and EQPO algorithms' property of terminating the trellis stages, when no Pareto optimal routes are detected. Thus, they are incapable of even examining potential Pareto-optimal routes located at later trellis stages. This limitation is partially mitigated by the SO-BTP, which rectifies the deficiency, where a globally optimal route may be located several stages apart from the rest of the OPF. Despite this inability, the BTA-EQPO algorithm's performance is near-optimal, identifying the Pareto optimal routes with 0.1\% probability of misdetection, while being able to detect 99.97\% of the time the true OPF.

\section{Conclusions}
We have further developed the quantum-assisted multi-objective dynamic programming framework of \cite{alanis2017eqpo} by introducing the SO-BTP and the MO-BTP for the sake of enhancing the heuristic accuracy attained. We have shown that the SO-BTP enables the algorithm to detect almost all of the Pareto optimal solutions, while the activation of MO-BTP also increases our confidence in detecting only the true Pareto-optimal routes. Finally, we have proven that the SO-BTP's extra complexity is insignificant. Furthermore, we have demonstrated for the MO-BTP that its extra complexity is insignificant as long as we employ less than 5 backward-trellis steps. Finally, we have demonstrated that with the above proviso the BTA-EQPO algorithm outperforms the EQPO and exhibits a near-optimal accuracy.

\bibliographystyle{ieeetr} 
\bibliography{mybib}

\end{document}